\documentclass[runningheads]{cl2emult}

\usepackage{epsfig}
\usepackage{makeidx}  
\usepackage{graphicx} 
\usepackage{subeqnar} 
\usepackage{multicol} 
\usepackage{cropmark} 
\usepackage{physbb}   
\makeindex            

 \newcommand{\beq}{\begin{equation}}
 \newcommand{\eeq}{\end{equation}}
 \newcommand{\bea}{\begin{eqnarray}}
 \newcommand{\eea}{\end{eqnarray}}

 \def\lr{\left( }
 \def\rr{\right) }
 \def\le{\left[ }
 \def\re{\right] }

\begin{document}
\title*{Photon Structure and the Production of Jets, Hadrons, and Prompt
        Photons}
\toctitle{Photon Structure and the Production of Jets, Hadrons, and Prompt
          Photons}
%
%
\titlerunning{Photon Structure and the Production of Jets, Hadrons, and Prompt
              \dots}
%
\author{Michael Klasen}
\authorrunning{Michael Klasen}
\institute{HEP Theory Group, Argonne National Laboratory,
           Argonne, IL 60439, USA}

\maketitle              

\begin{abstract}
We give a pedagogical introduction to hard photoproduction processes at HERA,
including the production of jets, hadrons, and prompt photons. Recent
theoretical developments in the three areas are reviewed.
\end{abstract}

\vspace*{-7.3cm} \noindent ANL-HEP-CP-99-73
\vspace*{ 6.4cm}

\section{Introduction}

Electron-proton scattering at HERA proceeds dominantly through the exchange of
a single photon with small virtuality $Q\simeq 0$. In this photoproduction
limit and in the presence of a hard factorization scale $M_{\gamma,p}$, the
electron-proton scattering cross section can be decomposed into
\beq
 \mbox{d}\sigma_{ep} =
 \sum_{a,b}\int\limits_0^1\mbox{d}yF_{\gamma/e}(y,Q_{\max}^2)
 \int\limits_0^1\mbox{d}x_{\gamma}F_{a/\gamma}(x_{\gamma},M_{\gamma}^2)
 \int\limits_0^1\mbox{d}x_pF_{b/p}(x_p,M_p^2)
 \mbox{d}\sigma_{ab}^{(n)}.
\eeq

The Weizs\"acker-Williams spectrum of photons in the electron
\beq
 F_{\gamma/e}(y,Q_{\max}^2) = \frac{\alpha}{2\pi} \le \frac{1+(1-y)^2}{y}
 \ln\left(\frac{Q_{\max}^2}{Q_{\min}^2}\right)
 +2m_e^2y\lr\frac{1}{Q_{\max}^2}-\frac{1}{Q_{\min}^2}\rr\re
\eeq
is proportional to the electromagnetic coupling constant $\alpha$. Since the
photon virtuality is small, the spectrum can be calculated explicitly by
exploiting current conservation and integrating over the unobserved azimuthal
angle of the outgoing electron and over the virtuality of the photon.
$Q_{\min}^2 = m_e^2 y^2/(1-y)$ depends on the electron mass $m_e$ and the
longitudinal momentum fraction $y$ of the photon in the electron. $Q_{\max}^2=
E_e^2 (1-y) \theta^2$ is determined experimentally from the incoming electron
beam energy $E_e$, the momentum fraction $y$, and the scattering angle of the
outgoing electron $\theta$.

From deep inelastic scattering (DIS) experiments of virtual photons off protons
it is well known that the proton has point-like constituents, quarks $q$ and
gluons $g$, also called partons. As the photon virtuality decreases, the photon
itself begins to fluctuate into quark-antiquark $(q\overline{q})$ pairs, which
in turn evolve into a vector meson-like structure. At HERA, an almost real
photon radiated from the electron can thus interact either directly with the
partons in the proton (direct component) or act as a hadronic source of partons
which collide with the partons in the proton (resolved component). In the
latter case, one does not test the proton structure alone but also the photon
structure. Unfortunately, both the photon and the proton structure cannot be
calculated theoretically, but have to be determined experimentally.

Before HERA started taking data, information on the parton densities in the
photon $F_{a/\gamma}$ came almost exclusively from deep inelastic
$\gamma^{\ast}\gamma$ scattering at $e^+e^-$ colliders. Whereas
the singlet quark densities can be
well constrained in this process, it is difficult to obtain information on the
gluon density, which is suppressed relative to the quark densities by the
strong coupling constant $\alpha_s$. The gluon density can, however, be
constrained from hard photoproduction processes at HERA if the direct
contribution is suppressed by measuring the production of jets, hadrons, or
prompt photons at low transverse energies $E_T$. This constitutes an important
goal in studies of hard photoproduction.

Among the parton densities in the proton $F_{b/p}$, the quark densities and
the gluon density at low $x_p$ are
rather well known today from DIS experiments. Of particular interest
in photoproduction is the gluon distribution in the proton at large $x_p$.
There, the experimental information from high-$E_T$ jet and prompt photon
production in hadron collisions is inconclusive and has considerable
potential for improvement from lepton pair production in hadron collisions and
from high-$E_T$ photoproduction processes. At large $E_T$, the direct
photoproduction process dominates and uncertainties from the photon structure
are suppressed.

In the presence of a hard renormalization scale $\mu \simeq E_T$, $\alpha_s
(\mu^2)$ becomes small and the partonic cross section
\beq
 \mbox{d}\sigma_{ab}^{(n)} = \alpha_s^0 (\mu^2) \mbox{d}\sigma_{ab}^{(0)}
                            +\alpha_s^1 (\mu^2) \mbox{d}\sigma_{ab}^{(1)}
                            +\dots
\eeq
for the scattering of two partons $a$ and $b$, which can be quarks, gluons,
and -- in the photon case -- also photons, into jets, hadrons, and prompt
photons can be calculated in perturbative quantum chromodynamics (QCD).
Testing these predictions constitutes a third goal for photoproduction
experiments.

\begin{figure}
\begin{center}
\epsfig{file=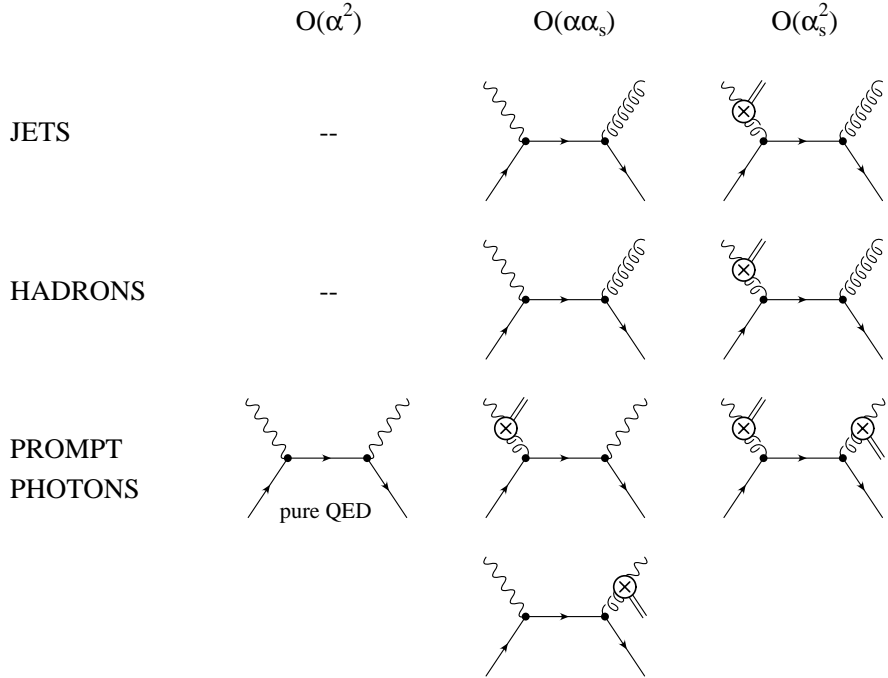,bbllx=34pt,bblly=43pt,bburx=435pt,bbury=352pt,%
 width=\textwidth}
\end{center}
\caption[]{Typical leading order QCD diagrams for the direct and resolved
 photoproduction of jets, hadrons, and prompt photons by order of the partonic
 scattering process. The ${\cal O}(\alpha^2)$ process is only present for the
 photoproduction of prompt photons}
\label{fig:1}
\end{figure}

Typical leading order (LO) QCD diagrams for the direct and resolved
photoproduction of jets are shown in the first line of Fig.\ \ref{fig:1}.
Next-to-leading order (NLO) QCD corrections to these graphs arise from virtual
loop and real emission diagrams and have been calculated in
\cite{Klasen:1998br}. The direct and resolved partonic processes are of
${\cal O}(\alpha\alpha_s)$ and ${\cal O}(\alpha_s^2)$, respectively. Since the
parton densities in the photon are of ${\cal O}(\alpha/\alpha_s)$ in the
asymptotic limit of large $M_{\gamma}^2$, both processes contribute at
${\cal O}(\alpha\alpha_s)$. In
addition to the goals of photoproduction studies mentioned above, the
observation of jets also permits to test and improve jet algorithms of both
the cone and cluster types and to study jet profiles and the internal jet
structure.

The LO partonic QCD diagrams for photoproduction of hadrons (second line in
Fig.\ \ref{fig:1}) are identical to those for the photoproduction of jets, and
the NLO corrections for inclusive hadron production have been calculated in
\cite{Kniehl:1994jk}. However, in this case the hadronic cross section
\beq
 \mbox{d}\sigma_{ab\rightarrow h}^{(n)} =
 \int\limits_0^1\frac{\mbox{d}z}{z^2} D_{c/h}(z,M_h^2)
 \mbox{d}\sigma_{ab\rightarrow c}^{(n)}
\eeq
is obtained
from the partonic cross section d$\sigma_{ab\rightarrow c}$ by
convoluting it with a fragmentation function $D_{c/h}$. Like parton
distribution functions, fragmentation functions cannot be derived from first
principles, and photoproduction of hadrons at HERA offers the additional
possibility to test hadron fragmentation functions obtained, {\it e.g.}, in
fits to data from $e^+e^-$ colliders.

Like initial state photons, final state (prompt) photons couple to the hard
partonic scattering process either directly or through their partonic
constituents. The direct (lower left diagram in Fig.\ \ref{fig:1}),
single-resolved (center), and double-resolved (right) prompt photon processes
contribute at different orders of the hard scattering process, but eventually
the photon structure and fragmentation functions compensate for these
differences so that all three subprocess types are effectively of ${\cal O}
(\alpha^2)$. Comparable to hadron fragmentation functions, the partonic content
of final state photons is described by a photon fragmentation function
$D_{c/\gamma}$. Photoproduction of isolated photons and photons in association
with jets has been calculated in NLO QCD in \cite{Gordon:1998yt}. Since prompt
photon production is a purely electromagnetic process in LO, its cross section
is smaller than the inclusive jet and hadron cross sections, and
data from H1 and
ZEUS have only recently become available.

\section{Jets}

At the first Ringberg workshop on ``New Trends in HERA Physics'' in 1997,
photoproduction of jets had already been studied for five years. Several NLO
single jet and the first NLO dijet calculations had been compared to data from
H1 and ZEUS \cite{Klasen:1997tq}. Studies
of jet cross sections \cite{Butterworth:1996ey} and jet shapes
\cite{Klasen:1997tj} had demonstrated that the commonly used Snowmass jet cone
algorithm \cite{Huth:1990mi} suffered from theoretical (double counting, parton
distance) and experimental (seed finding, jet overlap) ambiguities, which could
only be resolved by the introduction of a phenomenological parton distance
parameter $R_{\rm sep}$ in addition to the cone size parameter $R$.

These ambiguities were found to be absent in the $k_T$-clustering algorithm in
the inclusive mode \cite{Ellis:1993tq}, which has been used by the HERA
experiments since then. It combines two hadronic clusters $i$ and $j$ if their
distance
\begin{equation}
  d_{ij} = \min(E_{T,i}^2,E_{T,j}^2)[(\eta_i-\eta_j)^2+(\phi_i-\phi_j)^2]/R^2
\end{equation}
is smaller than $E_T^2$. The parameter $R=1$ now corresponds to the
parton distance parameter $R_{\rm sep}$ in NLO QCD.
As in the cone algorithm, the transverse energy $E_T$, rapidity
$\eta$, and azimuthal angle $\phi$ of the combined cluster are calculated from
the transverse energy weighted sums of the two pre-clusters.

\begin{figure}
\begin{center}
\epsfig{file=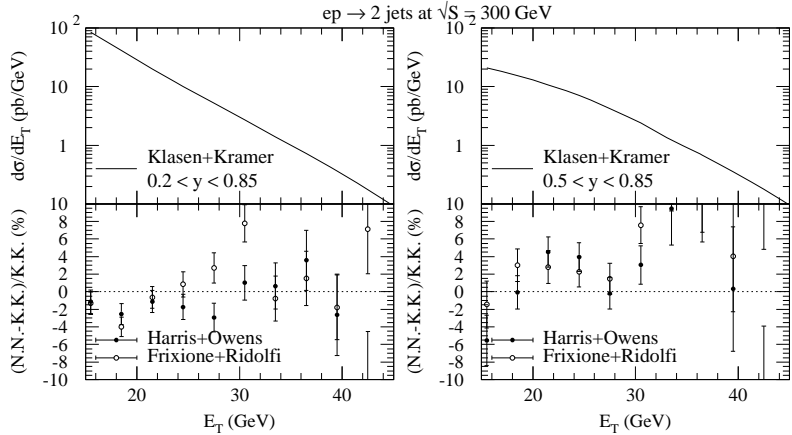,width=\textwidth}
\end{center}
\caption[]{Comparison of three theoretical predictions for the NLO dijet
 cross section as a function of the transverse energy $E_T$ of the leading
 jet for the full (left) and high (right) $y$ range. Both jets lie in a central
 rapidity range $0 < \eta_{1,2} < 1$}
\label{fig:2}
\end{figure}

Three different NLO dijet calculations have recently been compared to ZEUS
dijet data (see contribution by L.~Sinclair
to these proceedings) \cite{Breitweg:1999wi}. The definition of the ZEUS dijet
cross section has also served as a basis to compare the theoretical predictions
among themselves \cite{Harris:1998ss}. As can be seen in Fig.\ \ref{fig:2},
the NLO transverse energy distributions agree with each other within the
statistical accuracy of the Monte Carlo integration, which is about $\pm 1$\%
at low $E_T$ for the full $y$ range and $\pm 2$\% for the high $y$ range.

\begin{figure}
\begin{center}
\epsfig{file=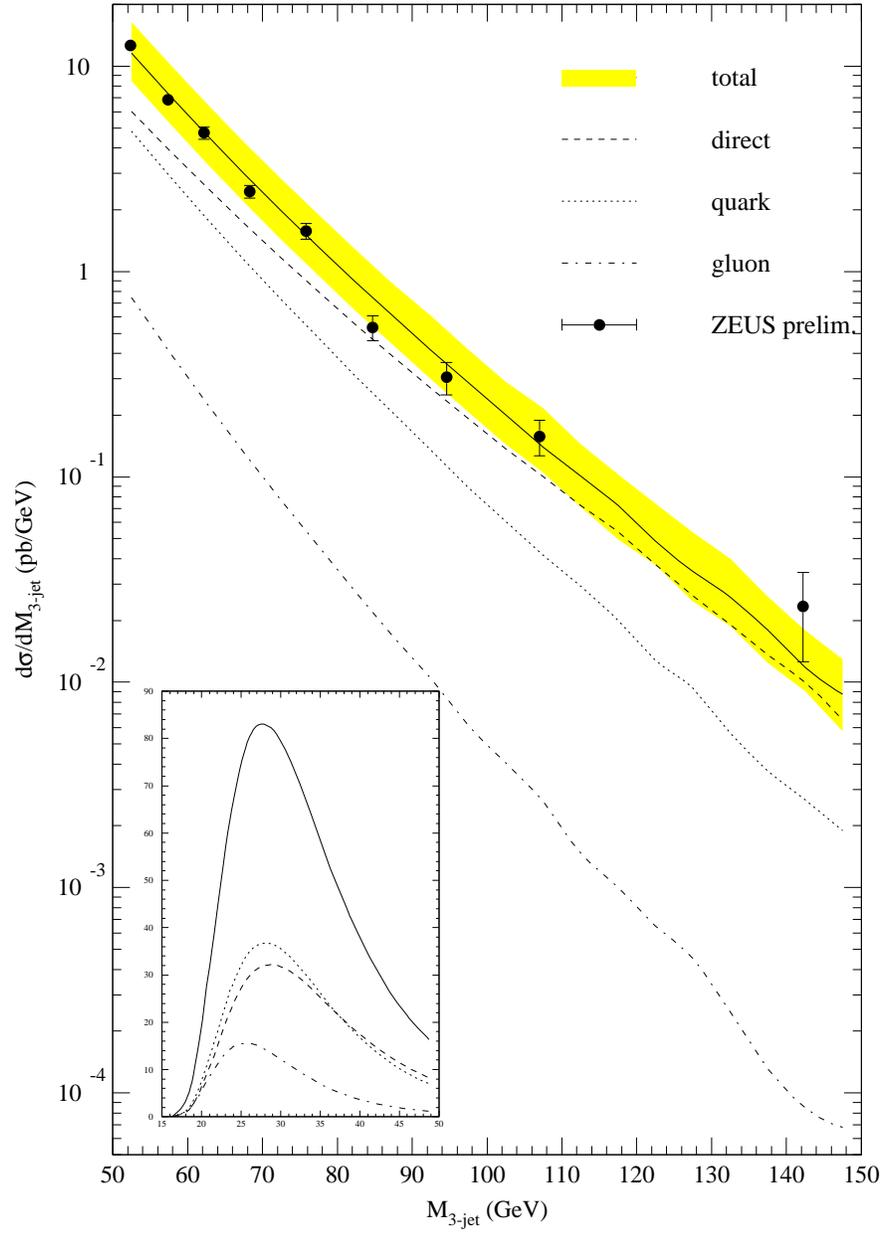,bbllx=60pt,bblly=95pt,bburx=495pt,bbury=715pt,%
 width=\textwidth}
\end{center}
\caption[]{Total cross section (full curve) for the photoproduction of three
 jets as a function of the three jet mass $M_{\rm 3-jet}$. We also show the
 variation of the absolute normalization due to the uncertainty in the scale
 choice (shaded band) and the contributions from direct photons (dashed),
 quarks (dotted), and gluons (dot-dashed) in the photon. The ZEUS data
 \cite{Breitweg:1998uv} agree well with the QCD prediction
 \cite{Klasen:1998cw}}
\label{fig:3}
\end{figure}

NLO dijet calculations involve the calculation of one-loop $2\rightarrow 2$ and
tree-level $2\rightarrow 3$ scattering matrix elements. The latter are then
integrated over soft and collinear regions of phase space in order to cancel
the infrared divergences arising from the virtual corrections. If used in their
original, unintegrated form, the $2\rightarrow 3$ matrix elements can also be
used for LO predictions of three jet cross sections. The three NLO
dijet predictions mentioned above have also been applied to LO three jet cross
sections and found to agree with each other \cite{Harris:1998ss}. ZEUS
have measured the photoproduction of three jets with transverse energies larger
than 6 GeV (two highest jets) and 5 GeV (third jet) and rapidities
$|\eta| < 2.4$ \cite{Breitweg:1998uv}. Fig.\
\ref{fig:3}
shows that the total theoretical prediction (full curve) describes the data
well in shape and normalization \cite{Klasen:1998cw}.
The agreement in normalization is, however, to
some degree coincidental since there is still an uncertainty of a factor of two
from the variation of the scales $\mu=M_{\gamma}=M_p\in [0.5;2.0] \times
\max(E_{T,1},E_{T,2},E_{T,3})$ (shaded band).
In these LO QCD predictions every parton corresponds to a
jet, but a jet definition still has to be implemented for the condition that
two partons cannot come closer too each other than permitted by the jet
algorithm.

\begin{figure}
\begin{center}
\epsfig{file=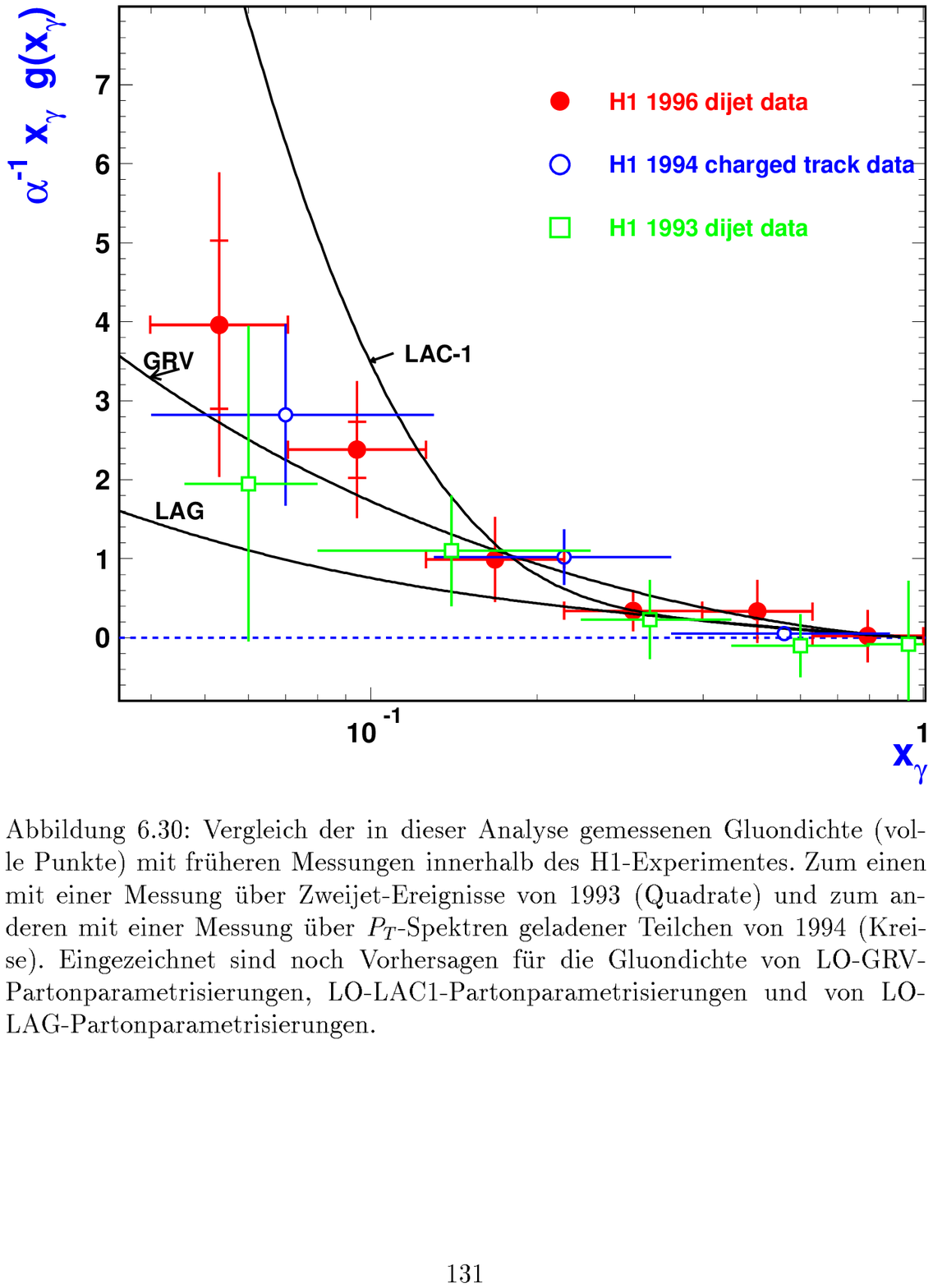,bbllx=85pt,bblly=305pt,bburx=499pt,%
 bbury=655pt,width=\textwidth,clip=}
\end{center}
\caption[]{H1 extraction of the gluon density in the photon from 1996 dijet
data \cite{Cvach:1999uc}, 1994 charged track data \cite{Adloff:1998vt}, and
1993 dijet data \cite{Ahmed:1995wv}. The data are consistent with each other
and with LO GRV parametrization \cite{Gluck:1992jc}. The 1996 data rule out
both the LAC1 \cite{Abramowicz:1991yb} and LAG \cite{Abramowicz:1997tj} gluon
parametrizations}
\label{fig:4}
\end{figure}

H1 have recently analyzed the production of dijets with transverse momentum
$p_T > 4$ and 6 GeV, dijet mass $M_{\rm 2-jet} > 12$ GeV, rapidities $-0.5 <
\eta_{1,2} < 2.5, |\eta_1-\eta_2|<1$, and photon energy fraction $0.5 < y <
0.7$ as a function of the variable
\begin{equation}
 x_{\gamma}^{\rm jets} = \sum_{i=1}^{2} E_{T,i}e^{-\eta_i} / (y\sqrt{S}).
\end{equation}
This variable can be determined from the two observed jets and is, in LO QCD,
directly related to the momentum fraction of the parton in the photon. H1 have
used this measurement of the dijet cross section to extract the gluon
density in the photon with the result shown in Fig.\ \ref{fig:4}.
The 1996 data \cite{Cvach:1999uc} are consistent with older H1 charged track
\cite{Adloff:1998vt} and jet \cite{Ahmed:1995wv} measurements.
They agree very well with the GRV parametrization \cite{Gluck:1992jc}, but
rule out both the LAC1 \cite{Abramowicz:1991yb} and LAG parametrizations
\cite{Abramowicz:1997tj} of the gluon density. The systematic (outer) error
bars include theoretical uncertainties coming from the variation of
different Monte Carlo models, parton densities in the proton, and quark
densities in the photon.

\section{Hadrons}

The transformation of quarks and gluons into hadronic
final states can be described globally by jet definitions. More detailed
experimental information about the hadronization process can be obtained in
the production of single hadrons, which is described theoretically by
fragmentation functions. Prior to the last Ringberg HERA workshop, several new
NLO QCD fits of fragmentations functions for charged pions, charged kaons, and
neutral kaons to $e^+e^-$ data from TPC, ALEPH, and Mark II had been performed.
They had been applied to a NLO QCD calculation for photoproduction and
successfully been compared to HERA data \cite{Kniehl:1997dr}. Since then,
fragmentation functions for heavy particles, such as $D^{*\pm}$ and $B$ mesons,
have been fitted to ALEPH and OPAL data, and they also compare favorably to
HERA data (see contribution by G.~Kramer to these proceedings). A new fit of
charged pion, charged kaon, and proton fragmentation functions, separately for
light flavors, heavy flavors, and gluons, to new LEP and SLC data is currently
in progress \cite{KKP}.

\begin{figure}
\begin{center}
\epsfig{file=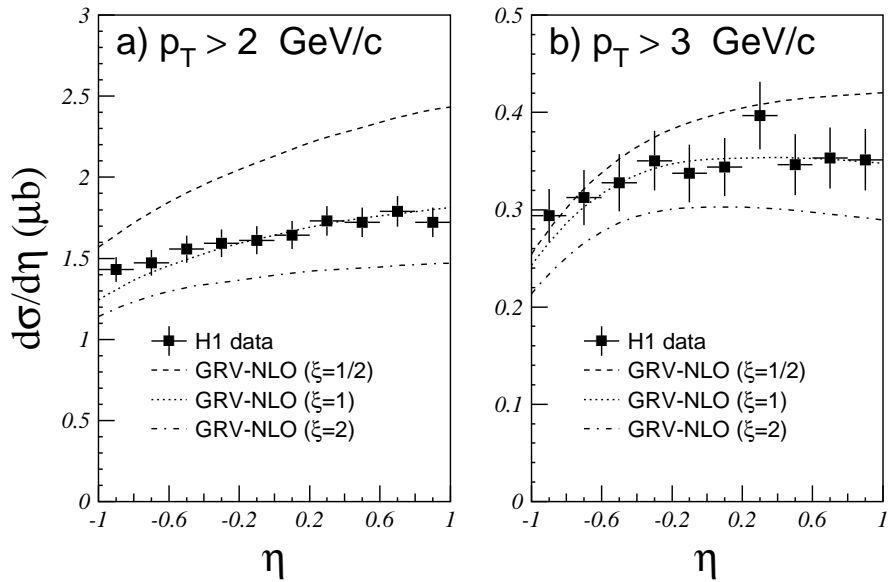,bbllx=64pt,bblly=278pt,bburx=527pt,bbury=583pt,
 width=\textwidth,clip=}
\end{center}
\caption[]{Differential cross section for the photoproduction of charged
hadrons as a function of rapidity for two different transverse momentum cuts.
Recent H1 data \cite{Adloff:1998vt} are compared to NLO QCD predictions
\cite{Binnewies:1995pt} with three different scale choices}
\label{fig:5}
\end{figure}

In Fig.\ \ref{fig:5}
we compare NLO QCD predictions for the rapidity distribution of charged
hadrons \cite{Binnewies:1995pt} to recent H1 data \cite{Adloff:1998vt}. Good
agreement is found for the GRV parton densities in the photon
\cite{Gluck:1992jc} and the proton \cite{Gluck:1992ng} and a central scale
choice of $\mu=M_{\gamma}=M_p=p_T$, but the theoretical scale uncertainty in
NLO QCD is still considerable. It is worth noting that in inclusive hadron
production, perturbative QCD works remarkably well down to very low transverse
momenta of $p_T > 2$ or 3 GeV. There is no excess in the forward $\eta$ region
as observed in jet photoproduction. Thus, inclusive hadron production offers
the potential to extract the gluon density in the photon in low-$p_T$ (and
therefore low-$x_{\gamma}$) cross section measurements.

\begin{figure}
\begin{center}
\epsfig{file=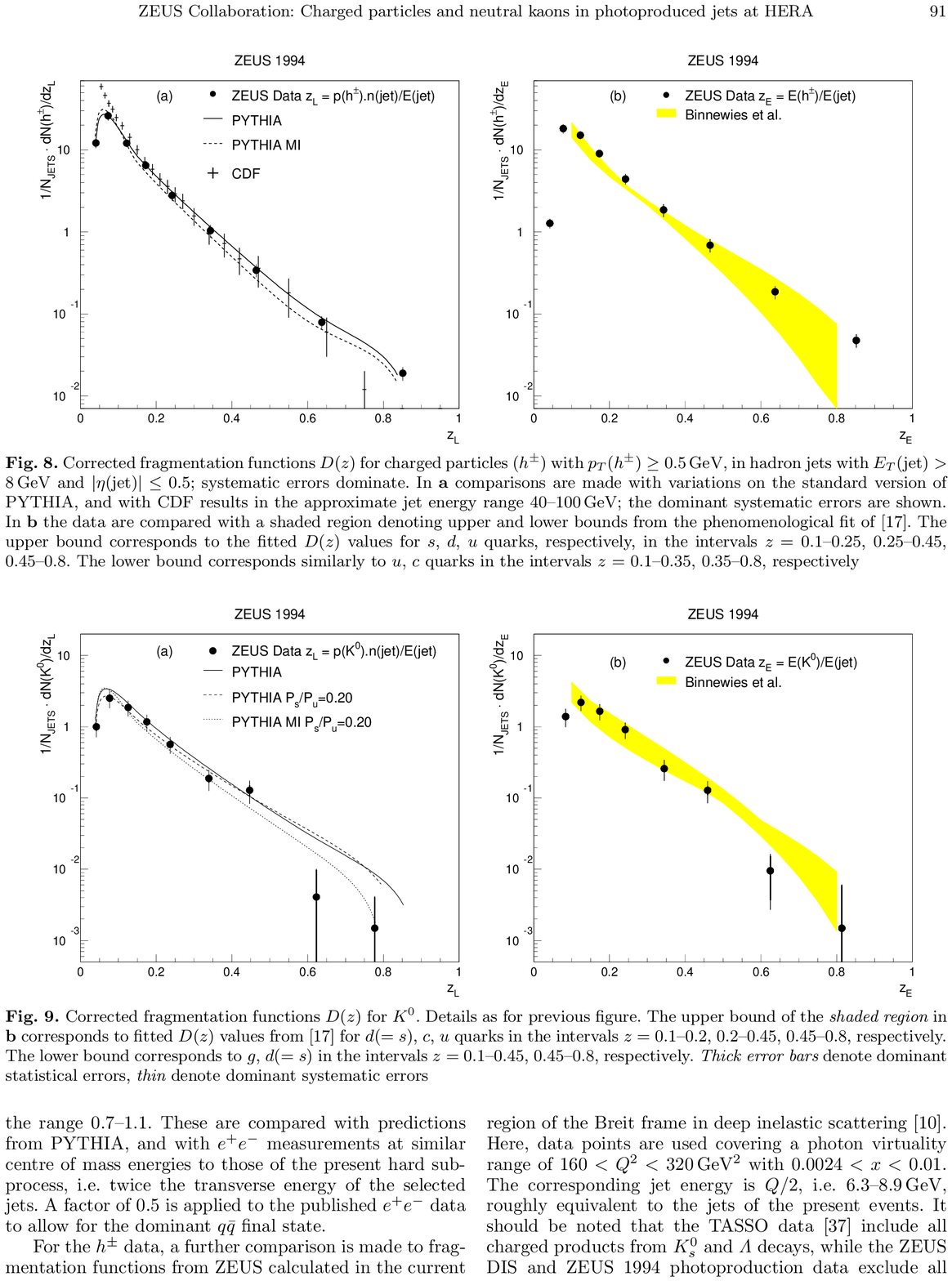,bbllx=302pt,bblly=531pt,bburx=536pt,%
 bbury=745pt,width=\textwidth,clip=}
\end{center}
\caption[]{Fragmentation function of charged particles as a function of their
longitudinal momentum fraction. The ZEUS data \cite{Breitweg:1997kc} for
particles within photoproduced jets are compared to a NLO fit to $e^+e^-$ data
\cite{Binnewies:1995pt}}
\label{fig:6}
\end{figure}

It is also interesting to check the universality of fragmentation functions
and the factorization theorem in photoproduction experiments. In Fig.\
\ref{fig:6}
we show a recent ZEUS fragmentation function measurement for charged hadrons
\cite{Breitweg:1997kc}, measured in photoproduced jets with $E_T>8$ GeV,
$|\eta|<0.5$, and $R=1$, which agrees nicely with the NLO QCD fit to
$e^+e^-$ data \cite{Binnewies:1995pt}. The error band in the NLO QCD fit
comes from a variation of the fragmenting quark flavor, which is unknown in the
ZEUS analysis.

\section{Prompt Photons}

\begin{figure}
\begin{center}
\epsfig{file=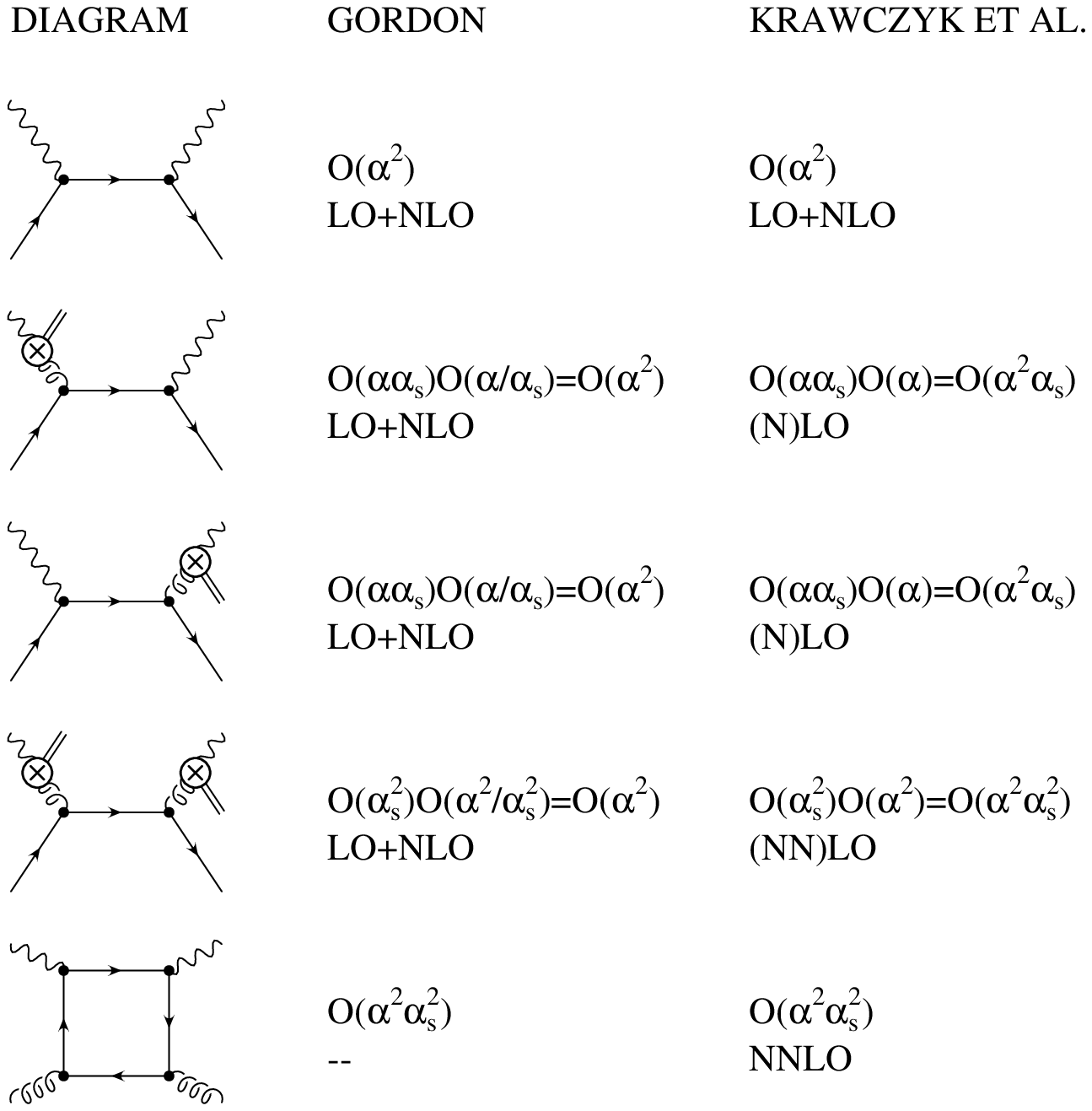,bbllx=70pt,bblly=42pt,bburx=440pt,bbury=421pt,%
 width=\textwidth}
\end{center}
\caption[]{Diagrammatic comparison of two prompt photon QCD calculations.
Gordon calculates the NLO corrections to the direct, single-resolved, and
double-resolved contributions \cite{Gordon:1998yt}, whereas Krawczyk {\it
et al.} only calculate the corrections to the direct contribution
\cite{Krawczyk:1998it}. On the other hand, Krawczyk {\it et al.} take into
account the box diagram that arises at NNLO}
\label{fig:7}
\end{figure}

\begin{figure}
\begin{center}
\epsfig{file=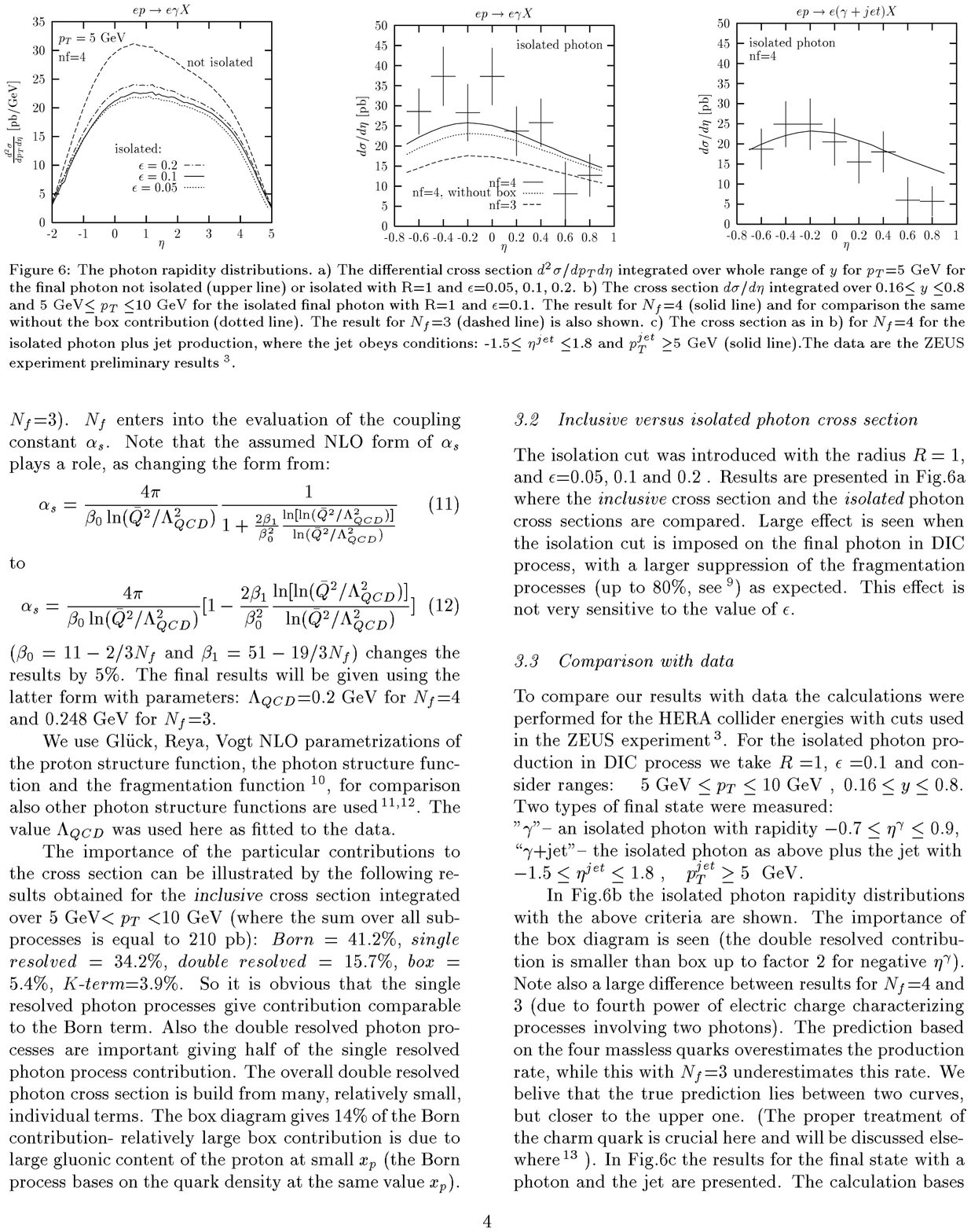,bbllx=241pt,bblly=577pt,bburx=384pt,%
 bbury=717pt,width=\textwidth,clip=}
\end{center}
\caption[]{Differential cross section for the photoproduction of a photon of
transverse energy $E_T\in[5;10]$ GeV, which is isolated in a cone with radius
$R=1$ and hadronic transverse energy fraction $\epsilon=0.1$, as a function of
rapidity. The ZEUS data \cite{Sinclair:1998hz} are compared to QCD predictions
with and without the NNLO box diagram and with three and four quark flavors
\cite{Krawczyk:1998it}}
\label{fig:8}
\end{figure}

Prompt photon production was not discussed at the 1997 Ringberg HERA workshop
-- it had yet to be observed. Only recently have H1 and ZEUS reported results
on this process \cite{Breitweg:1997pa,Muller:1997an}
and have two NLO QCD calculations for isolated photon
production and for the production of a photon in association with a jet become
available \cite{Gordon:1998yt,Krawczyk:1998it}. These two calculations differ
in their powercounting of the strong coupling constant and in their
consideration of higher-order corrections. To illustrate this, we have listed
in Fig.\ \ref{fig:7}
the direct, single-resolved, and double-resolved diagrams for photoproduction
of prompt photons. The single-resolved process includes either photon structure
or photon fragmentation contributions, whereas in the double-resolved case the
initial and final state photons contribute both through their partonic
constituents. Gordon counts the photon structure and fragmentation
functions as ${\cal O} (\alpha/\alpha_s)$ from considering the asymptotic
limit and calculates the NLO corrections to all three types of subprocesses.
Krawczyk {\it et al.} count the photon structure as ${\cal O} (\alpha)$,
calculate the NLO corrections only to the
direct process, and in addition take into account the box diagram in the last
line of Fig.\ \ref{fig:7}.
Although the box diagram is formally of NNLO, it
is known to have a large numerical contribution. This is demonstrated in Fig.\
\ref{fig:8}
where we compare the calculation by Krawczyk {\it et al.} using GRV photon
\cite{Gluck:1992jc} and proton \cite{Gluck:1992ng} structure and photon
fragmentation functions \cite{Gluck:1993zx} to ZEUS data for the
production of isolated photons with $E_T>5$ GeV \cite{Sinclair:1998hz}. With
higher luminosity and the ensuing better accuracy of the data, prompt photon
production at HERA will eventually provide useful tests of photon
structure and fragmentation functions.

\section{Summary}

In summary, hard photoproduction processes can provide very useful information
on the hadronic structure of the photon, in particular on the gluon density,
which is complimentary to the information coming from deep inelastic
photon-photon scattering at electron-positron colliders. Among the different
hadronic final states, jets are most easily accessible experimentally and
phenomenologically. On the other hand, inclusive hadron production offers the
possibility to test the universality of hadron fragmentation functions and
measure the photon structure down to very low values of $p_T$ and $x_{\gamma}$.
Prompt photon production suffers from a reduced cross section and limited data,
but allows for the additional testing of photon fragmentation functions. \\

\noindent {\bf Acknowledgment.} I would like to thank the organizers of the
Ringberg workshop for the kind invitation. This work has been supported by the
U.S.\ Department of Energy under Contract W-31-109-ENG-38.

\clearpage
\addcontentsline{toc}{section}{Index}
\flushbottom
\printindex

\end{document}